\documentclass[pdflatex,twocolumn,sn-basic,Numbered]{sn-jnl}

\usepackage{graphicx}%
\usepackage{subfigure}
\usepackage{multirow}%
\usepackage{amsmath,amssymb,amsfonts}%
\usepackage{amsthm}%
\usepackage{mathrsfs}%
\usepackage[title]{appendix}%
\usepackage{xcolor}%
\usepackage{textcomp}%
\usepackage{manyfoot}%
\usepackage{booktabs}%
\usepackage{algorithm}%
\usepackage{algorithmicx}%
\usepackage{algpseudocode}%
\usepackage{listings}%

\theoremstyle{thmstyleone}%
\theoremstyle{thmstyletwo}%

\theoremstyle{thmstylethree}%

\begin{document}

\title[Article Title]{Scale Free Projections Arise from Bipartite Random Networks}


\author*{\fnm{Josh} \sur{Johnston}}\email{jjohnston@u.boisestate.edu}

\author{\fnm{Tim} \sur{Andersen}}\email{tandersen@boisestate.edu}

\affil{\orgdiv{Department of Computer Science}, \orgname{Boise State University}, \orgaddress{\street{777 W Main St}, \city{Boise}, \postcode{83702}, \state{ID}, \country{USA}}}


\abstract{
The degree distribution  of a real world network --- the number of links per node --- often follows a power law, with some hubs having many more links than traditional graph generation methods predict. For years, preferential attachment and growth have been the proposed mechanisms that lead to these scale free networks. However, the two sides of bipartite graphs like collaboration networks are usually not scale free, and are therefore not well-explained by these processes.  Here we develop a bipartite extension to the Randomly Stopped Linking Model and show that mixtures of geometric distributions lead to power laws according to a Central Limit Theorem for distributions with high variance. The two halves of the actor-movie network are not scale free and can be represented by just $5$ geometric distributions, but they combine to form a scale free actor-actor unipartite projection without preferential attachment or growth. This result supports our claim that scale free networks are the natural result of many Bernoulli trials with high variance of which preferential attachment and growth are only one example.
}




\maketitle

\section{Introduction and Background}\label{sec1}

For two decades, the Barab\'asi-Albert (BA) Model has explained why power laws and other heavy-tailed distributions often emerge in what are known as scale free networks \cite{barabasi_2009}.  They propose that degree distribution in a real network --- the number of links per node --- tends toward a power law due to preferential attachment \cite{albert_et_al_1999}.  Recent work showed that preferential attachment and growth are not required to generate scale free networks \cite{johnston_andersen_2020}.  Linking processes behaving as Bernoulii trials with high variance also result in power law degree distributions.  This is predicted by a Central Limit Theorem (CLT) that says mixtures of geometric distributions with high variance will follow a power law \cite{johnston_andersen_2020}. The critical element of scale free networks are high variance linking probabilities, not preferential attachment per se.  We can build synthetic networks with the Randomly Stopped Linking Model \cite{johnston_andersen_2022}, which uses mixtures of geometric distributions to model Bernoulli processes with high variance and then links nodes together with a reparameterization of the Configuration Model \cite{bollobas_1980}.

In this paper, we extend the Randomly Stopped Linking Model to bipartite graphs.  These graphs have two types of nodes where links are only made from one type to the other \cite{guillaume_latapy_2006}.  Analyzing the actor-movie network provides insight to how projections of bipartite graphs can have power law degree distributions even when the distribution of each half of the network does not follow a power law.  We use a reparameterized Bipartite Configuration Model to reconfigure the links between actors and movies, demonstrating that preferential attachment is not needed to result in a projected actor-actor network similar to the real world version. We further show that a synthetic network created with geometric (non-heavy-tailed) distributions produces a power law degree distribution in the projected actor-actor network.

\subsection{The Actor-Actor Network}

The network that links Hollywood actors who have appeared in movies together is one example of a scale free network \cite{wasserman_faust_1994}, \cite{watts_strogatz_1998}.  Like other co-occurrence graphs such as identity networks and scientific collaboration networks, the actor-actor network is actually a projection from a bipartite network to a unipartite network.  In the bipartite newtork, a link connects from an actor node to a movie node to show an actor has appeared in a movie \cite{barabasi_marton_2017}. The version used to support the BA model is a projection where links connect actors directly to each other if they have been in a movie together \cite{barabasi_albert_1999}.  This paper uses the terminology ``actor-movie network'' for the full bipartite network and ``actor-actor network'' for the projected view.  

\subsection{Fitting a Power Law to Degree Distribution}

To consider a network scale free, we expect a power law to be a better fit than other distributions across at least 2--3 orders of magnitude in the degree distribution.  We use a method \cite{alstott_bullmore_plenz_2014} comparing the fit of a power law to a stretched exponential, using maximum-likelihood estimation (MLE) to determine whether a power law fit is best according to the Bayesian Information Criterion (BIC).  Even in scale free degree distributions, power laws are rarely the best distribution to fit all of the data, and we are most interested in the tail where power laws tend to become visible \cite{clauset_shalizi_newman_2009}.  So as long as the MLE does not support a power law fit, the method iteratively increases $k_{min}$ and tests the fit for data greater than this degree threshold.

\begin{figure}[]
	\centering
	\begin{subfigure}[]
		\centering
		\includegraphics[width=\columnwidth]{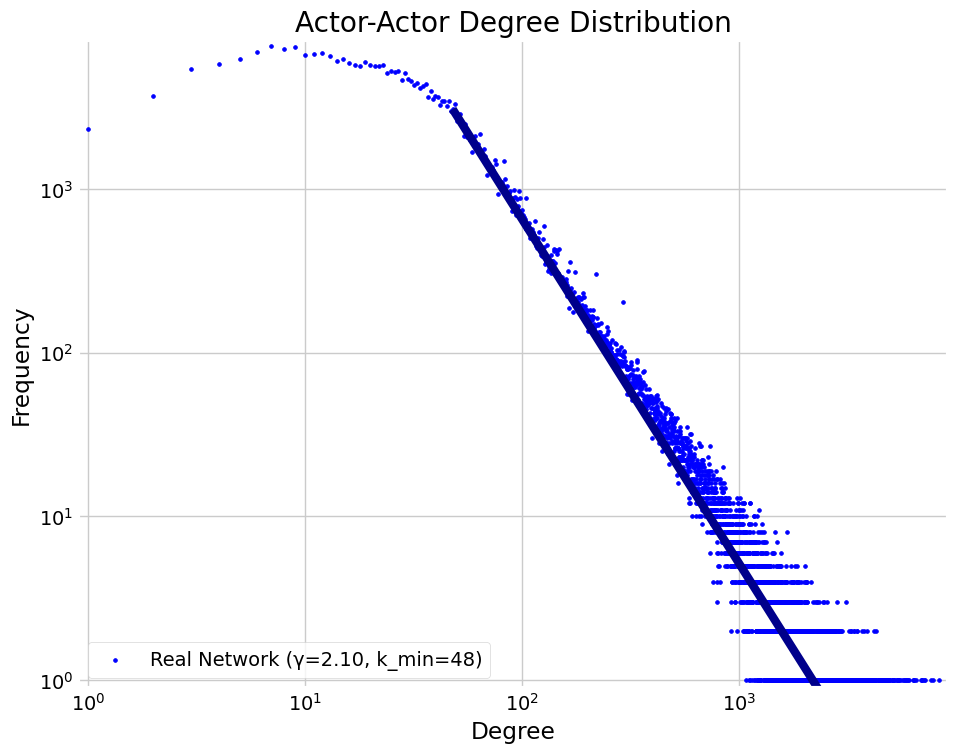}
		\label{fig:actor_actor}
	\end{subfigure}%
	\begin{subfigure}[]
		\centering
		\includegraphics[width=\columnwidth]{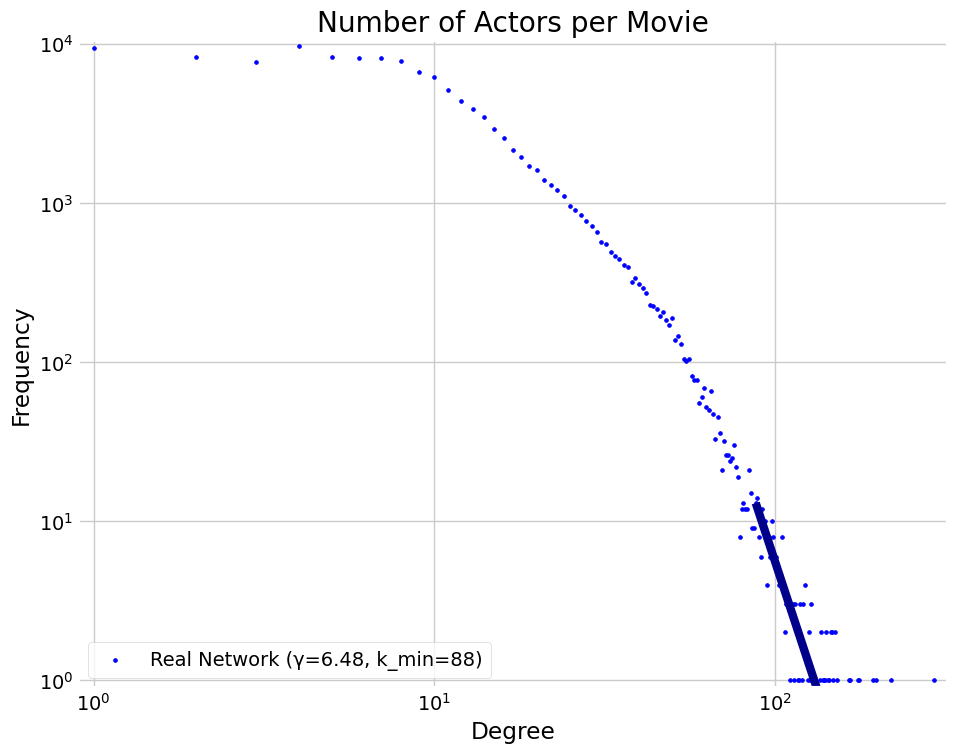}
		\label{fig:movie_nodes}
	\end{subfigure}%
	\begin{subfigure}[]
		\centering
		\includegraphics[width=\columnwidth]{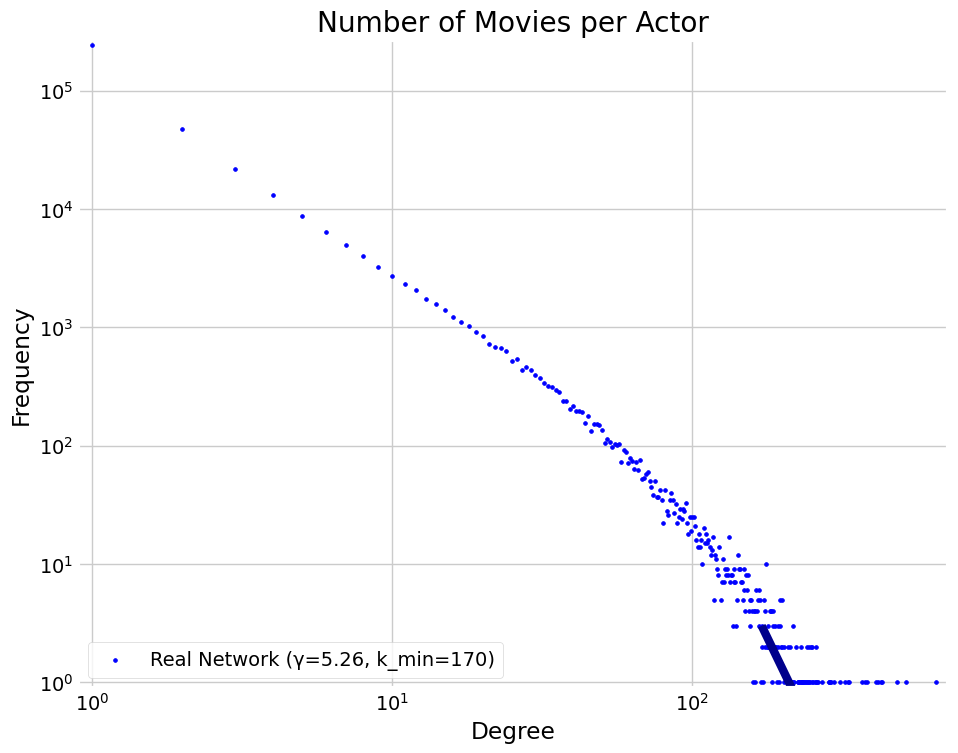}
		\label{fig:actor_nodes}
	\end{subfigure}%
	\caption{Three views of the Hollywood actor-movie network used for this paper, with data from \cite{pajek_2004}.  (a) shows the degree of actor nodes when the bipartite graph is projected into an actor-actor view, with links connecting actors who appeared in the same movie together.  The degree of the movie nodes (b) and actor nodes (c) from the underlying bipartite graph are also the number of actors in each movie (b) and the number of movies each actor appears in (c). $\gamma$ is the exponent of the fitted power law and $k_{min}$ is the minimum value for which the power law is the best fit to the data. }\label{fig:real_network}
\end{figure}

As seen in Figure \ref{fig:real_network}, the degree distribution of the projected actor-actor network follows a power law more closely than either of the actor or movie degree distributions in the bipartite network.  As we show below, the number of actors per movie is in fact fit well by a geometric distribution and is not heavy-tailed, much less scale free.  It has been previously noted that the two halves of collaboration networks, including the actor-movie network, have degree distributions that are usually not as heavy-tailed as the projected unipartite network \cite{guillaume_latapy_2006}.

\subsection{How the Actor-Movie Network is Formed}

The BA Model assumes new actors preferentially appear in movies with high-degree actors because ``a new actor is most likely to be cast in a supporting role with more established and better-known actors'' \cite{barabasi_albert_1999}.  This is an unproven assertion with plausible counterclaims, ie `new actors are more likely to appear in movies with other unknown actors than with a big star'.  The bipartite network is a more complete representation of the relevant relationship dynamics than the projected view \cite{newman_watts_strogatz_2002}.  Actor-actor links do not typically form as organic collaborations of actors.  Instead, movies are distinct entities, each with a fixed number of roles.  Actors compete for these roles.  While it makes sense that landing a well-respected actor will influence further casting, this potential network effect is likely small relative to the prior fitness of each actor competing for the role.  

In any event, our model ignores any network effects and assumes actor-actor degree is dominated by the number of movies an actor appears in and the number of actors per movie, without regard to preferential attachment.  Therefore, we explore two questions separately: 
\begin{enumerate}
\item How many actors are in each movie?
\item How many movies does each actor appear in?
\end{enumerate}
The answers to these questions become the degree distributions for each type of node.  First, we will combine the two halves from the original degree distributions using a Bipartite Configuration Model, showing that preferential attachment is not required in the node linking step to produce a scale free network.  Later, we will show that the degree distributions of each half of the bipartite model can be parameterized as geometric distributions and still result in a scale free projected actor-actor network.

\section{Relinking the actor-actor network using the Bipartite Configuration Model}

In the Bipartite Configuration Model, the two halves of the network (movie and actor) are initialized as link stubs attached to each node. At this point, the stubs do not yet connect to other nodes to form actual links. The number of stubs for each node could be drawn from a distribution, but in the first example we will use the original degree distribution of each half of the real actor-movie network. Conceptually, we break every link and will relink new pairs of nodes without regard to how pairs were linked in the real network.

The stubs are combined to form new links using a technique to build a random bipartite network from prescribed degree distributions\cite{guillaume_latapy_2006}].  We create the graph by randomly selecting pairs of unlinked stubs --- a movie stub and an actor stub --- then linking them together.  This step repeats until there are no more stubs \cite{tian_he_liu_du_2012}.

\begin{figure}[]
	\centering
	\includegraphics[width=\columnwidth]{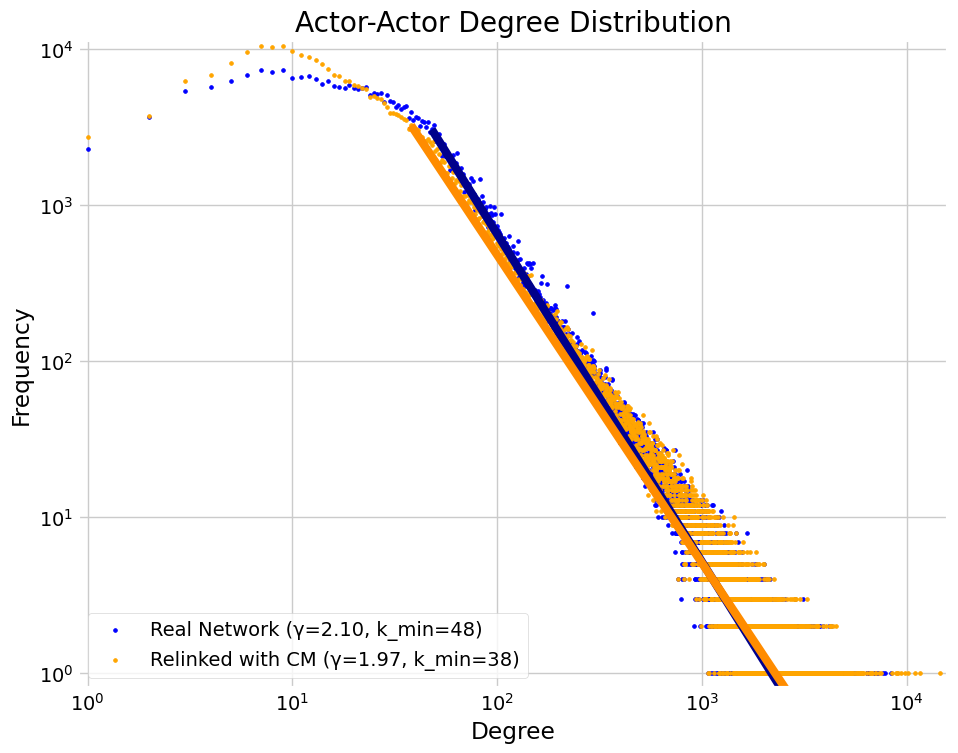}
	\caption{Comparison of the original, real network to one relinked via the Bipartite Configuration Model. Here, movies and actors are linked randomly without regard for preferential attachment according to the degree distribution of the movie and actor nodes from the original. The figure shows the degree distribution of the projected actor-actor network from the resulting bipartite network.}
	\label{fig:relinked_comparison}
\end{figure}

After creating the bipartite actor-movie network, we extract the actor-actor projection and compare its degree distribution with that of the real network.  Our model emphasizes two distinct, independent processes rather than being driven by preferential attachment or other network effects: movie writers create a number of roles according to one process and actors are cast in a number of movies according to another process.  Then they are linked randomly (Figure \ref{fig:relinked_comparison}).

\section{From relinking to a fully synthetic network}

We have shown that a scale free actor-actor network emerges from the two halves of a bipartite network when relinked randomly, without preferential attachment or other network effects. Since we used the original degree distributions for actor nodes and movie nodes, however, there may be network effects responsible for generating those distributions that explain the scale free nature of the relinked and projected network.  In this section, we use geometric distributions to parameterize the two halves of the bipartite network, again without preferential attachment or network effects.  This shows that Bernoulli processes can lead to scale free bipartite networks as predicted by the CLT for high variance distributions.

\subsection{How many actors are in each movie?}

\begin{figure}[]
	\centering
	\includegraphics[width=\columnwidth]{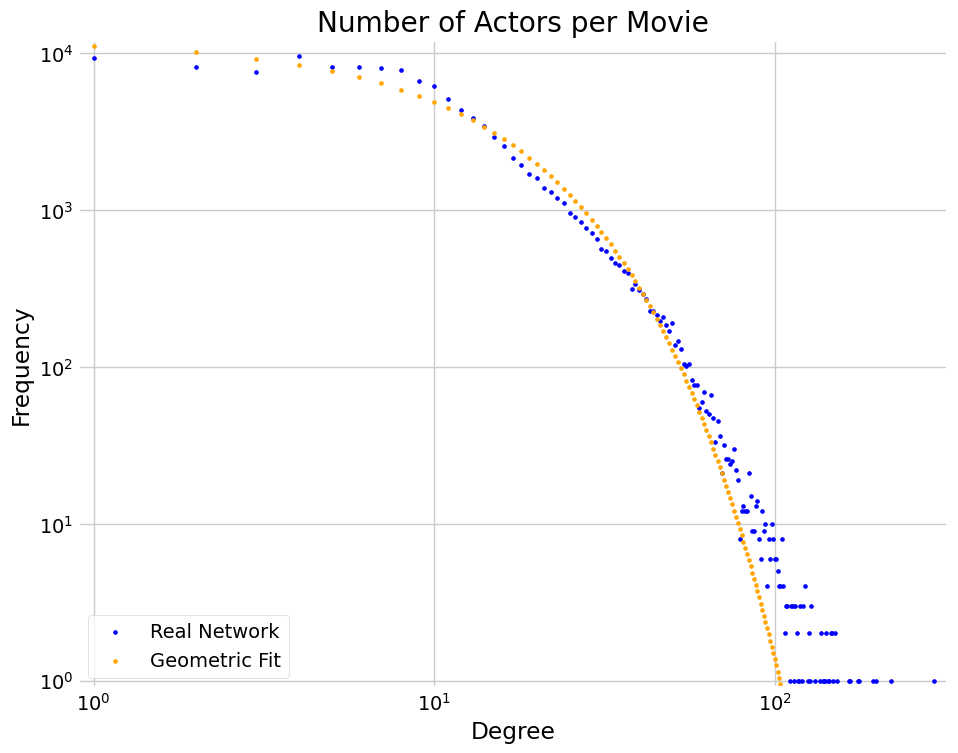}
	\caption{A geometric distribution fit to the movie node degree distribution. Given that $\mu = 11.5$ in the real network, $p = 0.087$ according to Equation \ref{eq:p}.}
	\label{fig:geometric_fit}
\end{figure}

In the bipartite actor-movie network, the number of actors in each movie is the degree of movie nodes (Figure \ref{fig:movie_nodes}).  Our model recognizes that each role added to a movie is a discrete decision made in series.  After the first role, there is a chance the writer will add another role.  If not, the process ends.  If the writer adds a second role, there is now a chance to add a third role, and so on.  In the simplest approximation, we consider the marginal probability of adding each role to be the same.  That process is described by a geometric distribution (Equation \ref{eq:geometric}), the number $k$ Bernoulli trial failures before the first success.  Equation \ref{eq:mean} finds $\mu$, the mean of the distribution, in terms of the parameter $p$, which is then rearranged as Equation \ref{eq:p}.  In the real network, $\mu = 11.5$, so according to Equation \ref{eq:p}, $p = 0.087$.  The result of this fit is presented as Figure \ref{fig:geometric_fit}.

\begin{equation}
	\label{eq:geometric}
	f(k) = (1-p)^k p
\end{equation}
\begin{equation}
	\label{eq:mean}
	\mu = \frac{1-p}{p}
\end{equation}
\begin{equation}
	\label{eq:p}
	p = \frac{1}{1+\mu}
\end{equation}

\subsection{How many movies does each actor appear in?}

Each actor in the network appears in some number of movies; this is the degree of the actor nodes in the actor-movie network.  Unlike movie degree, a single geometric distribution does not fit the actor node degree distribution very closely.  Instead, we use a mixture of geometric distributions, following the insight provided by the Randomly Stopped Linking Model \cite{johnston_andersen_2022}.  Expected outcomes for actors have high variance; not everyone starts with the same chance of making it big.  As a heterogeneous and constant property of nodes, a value of fitness can represent the competitive strength of each actor for roles \cite{bianconi_barabasi_2001}  This follows from, and is justified by, the observation that some actors have a priori advantages and therefore higher fitness for being cast in movies than others.  

\begin{figure}[]
	\centering
	\includegraphics[width=\columnwidth]{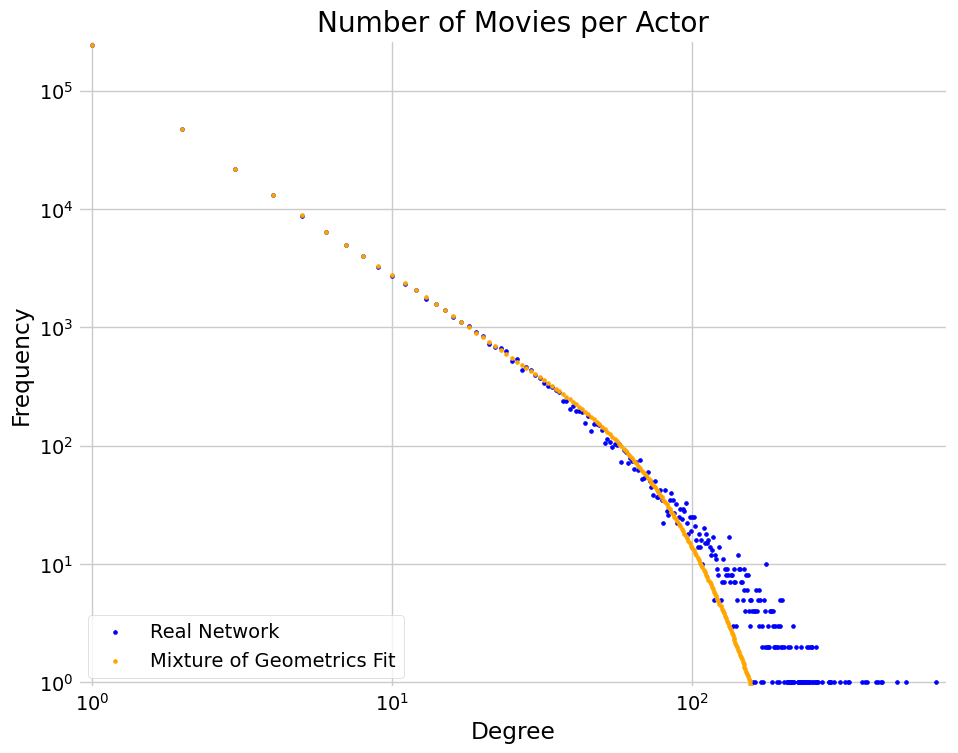}
	\caption{A mixture of geometric distributions fit for the actor node degree distribution.  The four geometric distributions use $p=\{0.046, 0.184, 0.528, 0.940\}$ and weight coefficient of $a=\{0.094, 0.178, 0.311, 0.562\}$, respectively.}
	\label{fig:mixture_geometrics_fit}
\end{figure}

Adapting the geometric distribution mixing function from\cite{johnston_andersen_2022}, we fit four geometric distributions to the actor node degree distribution.  This fit uses $8$ parameters, with each distribution having a value for $p$ in (Equation \ref{eq:p}) as well as a coefficient weight $a$, and is performed with the Trust Region Reflective technique implemented by the SciPy $curve\_fit$ function\cite{scipy} \cite{vugrin_et_al_2007}.  The values of $p$ and $a$ that best fit the real network are presented in Table \ref{tab:best_fit_parameters} and the result of this mixture is shown as Figure \ref{fig:mixture_geometrics_fit}.

We can interpret the values of $p$ as the chance a member of a cohort of actors has not been cast in another movie.  The corresponding value of $a$ is a description of the size of that cohort.  The fit tells us that a priori, many more actors are expected to be in a small number of movies than are expected to make it big.

\begin{table}[]
\caption{Parameters for the best fit of four geometric distributions to the number of movies per actor in the actor-movie network.}\label{tab:best_fit_parameters}
\begin{tabular}{c|c}
p & a\\
\hline
0.046 & 0.094 \\
0.184 & 0.178\\
0.528 & 0.311\\
0.940 &  0.562 \\
\end{tabular}
\end{table}

\subsection{Generating Synthetic Networks with the Bipartite Configuration Model}

We have characterized the degree distribution of each half of the bipartite actor-movie network using parameterized distributions: geometric for the movie node degree and a mixture of four geometrics for the actor node degree.

\begin{figure}[]
	\centering
	\includegraphics[width=\columnwidth]{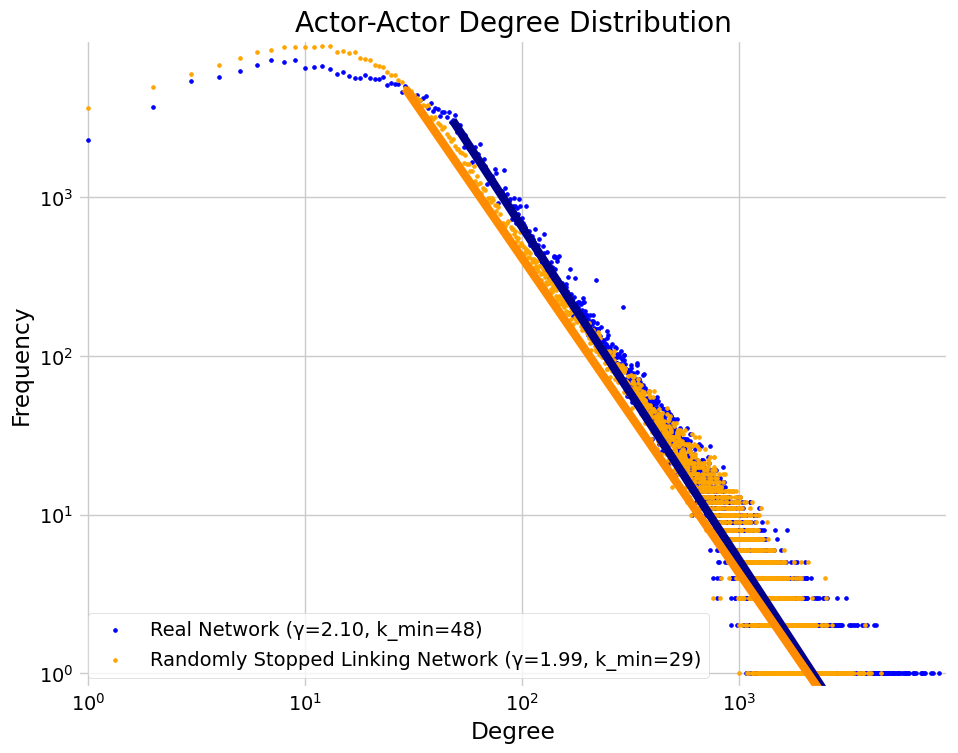}
	\caption{Comparison of the real actor-actor network to one created with the bipartite Randomly Stopped Linking Network and a total of $5$ different stopping probabilities represented by $5$ geometric distributions.}
	\label{fig:actor_actor_generated}
\end{figure}

To get from these PMFs to a network, we create the same number of movies and actors as contained in the real network.  Each generated movie is a node with a number of stub actor links pulled from a geometric distribution with the parameter fit earlier (Figure \ref{fig:geometric_fit}).  Separately, each actor is created as a node and assigned a number of movie roles by pulling from the PMF generated as a collection of geometric distributions (Figure \ref{fig:mixture_geometrics_fit}).  At this point, we have separately established the degree distributions for actors and movies, so each node has a certain number of stubs.  The modeled network is formed by randomly selecting a stub from an actor node and a stub from a movie node, then replacing those stubs with a link \cite{tian_he_liu_du_2012}.  As in the earlier relinking, there is no preferential attachment --- links connect without regard to how many actors each movie is already connected to, and vice versa.  The process repeats until all stubs are connected, creating the bipartite structure of the actor-movie network.  Finally, the actor-actor network is projected and compared to the real network (Figure \ref{fig:actor_actor_generated}). This synthetic network is the bipartite adaptation of the Randomly Stopped Linking Model from \cite{johnston_andersen_2022}.

The degree of our modeled actor-actor network is similar to that of the real network, but visual inspection on a log-log plot is insufficient to determine whether a power law is a better fit than other heavy-tailed distributions such as the log-normal \cite{stumpf_porter_2012}.  We use the two-sided Kolmogorov-Smirnov (KS) test to quantify goodness of fit \cite{clauset_shalizi_newman_2009}, then compare with a power law fit to the tail of the distribution above a $k_{min}$ \cite{scipy} \cite{vugrin_et_al_2007}.  A low KS statistic indicates a close fit between two distributions.

The KS statistic for the Randomly Stopped Linking network compared to a power law fit is $0.027$, which is near to, but worse than, the real network vs. power law KS stastistic of $0.022$. However, the $k_{min}$ of the Randomly Stopped Linking network is $29$ compared to the real network's $48$.  That means $46.6\%$ of the synthetic network's data is best described by a power law, compared to only $38.3\%$ of the real network.  This result shows that while the real network is different from the synthetic one built with geometric distributions, they are both scale free networks.  Additional statistics characterizing the real network, the geometric fit of the actor and movies nodes, and the synthetic Randomly Stopped Linking network are shown as Table \ref{tab:network_statistics}.

\begin{sidewaystable}[]
	\caption{Comparison of the real network and the Randomly Stopped Linking network, including the two halves of the bipartite network and the projected actor-actor view.}
	\label{tab:network_statistics}
	\begin{tabular}{l|l|l|l|l|l|l}
		& \multicolumn{3}{c|}{\textbf{Power Law Fit}} & \multicolumn{3}{c}{\textbf{Distribution Statistics}} \\
		\hline
		\textbf{Network} & \multicolumn{1}{c|}{\textbf{$\gamma$}} & \multicolumn{1}{c|}{\textbf{$k_{min}$}} & \multicolumn{1}{c|}{\textbf{Data Fraction}} & \multicolumn{1}{c|}{\textbf{\begin{tabular}[c|]{@{}c@{}}Variance\\ $\sigma^2$\end{tabular}}} & \multicolumn{1}{c|}{\textbf{\begin{tabular}[c]{@{}c@{}}Mean\\ $\mu$\end{tabular}}} & \multicolumn{1}{c}{\textbf{\begin{tabular}[c]{@{}c@{}}VMR\\ $\sigma^2/\mu$\end{tabular}}} \\
		\hline
		Real Network: Actors per Movie & $6.5$ & $88$ & $0.2\%$ & $138.2$ & $11.5$ & $12.0$ \\
		Geometric Fit: Actors per Movie & $4.5$ & $36$ & $4.2\%$ & $120.8$ & $11.5$ & $10.5$   \\
		Real Network: Movies per Actor  & $5.3$  & $170$ & $0.0\%$ & $108.7$ & $3.8$  & $28.4$ \\
		Mixture of Geometric Fit: Movies per Actor & $2.8$ & $31$ & $2.0\%$ & $72.5$   & $3.7$  & $19.8$ \\
		Real Network: Actor-Actor Degree & $2.1$ & $48$ & $38.3\%$ & $44514.9$ & $86.6$  & $513.8$\\
		Randomly Stopped Linking Network: Actor-Actor Degree & $2.0$ & $29$ & $46.6\%$ & $30059.0$ & $73.8$  & $407.2$                                                       
	\end{tabular}
\end{sidewaystable}

\section{Discussion}

\subsection{Heavy-Tailed Projections Emerge from Bipartite Networks without Heavy Tails}

Both the bipartite Randomly Stopped Linking network and the real network actor-actor degree distributions are well-characterized by power laws for several orders of magnitude, and can thus be considered scale free. The results in Table \ref{tab:network_statistics}, however, show that neither the actor nor movie components of the bipartite networks are scale free. The $k_{min}$ of $88$ for the Actors per Movie and $170$ for Movie per Actor means that a power law best describes only a trivial range of the degree distribution.  This is likely why most analyses showing collaboration networks to be scale free focus on the projection of the network.  Figure \ref{fig:actor_actor_generated} shows that the power law fit covers about $3.5$ orders of magnitude in this projected actor-actor network, which should therefore be considered scale free.

The result of projection is also visible in statistics from each network.  Both the variance and variance-to-mean ratio (VMR) are much higher after projection than in either half of the original bipartite graph.  This shows that a projection of a collaboration network can be scale free even when both halves and the original bipartite network are not. The two halves of the bipartite network have quite different degree distribution means and the variance increases substantially even when linked randomly without preferential attachment.  This is another example of high variance Bernoulli processes leading to scale free networks.

\subsection{Mixtures of High Variance Geometric Distributions Lead to Scale Free Networks}

Table \ref{tab:network_statistics} also compares the real actor collaboration networks with an example generated from the modeling with geometric distributions described earlier in this paper.  The degree distributions are indistinguishable between the generated and real networks.  The number of actors per movie is drawn from a single geometric distribution and the number of movies per actor is pulled from a mixture of four geometric distributions.  Therefore, the generated version of the projected actor-actor network is formed from only five geometric distributions, then linked according to the Configuration Model with no growth or preferential attachment.  

We have previously shown that the Randomly Stopped Linking Model creates scale free networks from Bernoulli trials with high variance \cite{johnston_andersen_2020}.  This paper extends the result by showing even a small number of geometric distributions can result in high enough variance to create a scale free network, especially when two halves of a bipartite network have widely separated means that increase the variance when assembled.

\section{Conclusions}

Since the discovery of power law degree distributions in real networks, preferential attachment has stood as the generally-assumed mechanism of their formation.  We demonstrate experimentally that independent Bernoulli Processes --- implemented by a small number of geometric distributions --- accurately model the actor-actor network's degree distribution without growth or preferential attachment.  Our technique has several advantages over the BA model, including:

\begin{enumerate}[1.]
	\item estimating the entire distribution rather than the minority of points in the right hand tail best fit by a power law
	\item recovering the full bipartite actor-movie graph rather than only the projected actor-actor view
	\item explaining the real network's distribution with a general theory of fitness variance rather than switching from a power law to a crossover distribution --- such as the stretched exponential --- in the presence of a nonlinear preferential attachment regime
\end{enumerate}
Abstracting from this specific actor-actor network, a generalized CLT provides a theoretical justification that heavy-tailed degree distributions are expected when the fitness of nodes has high variance.  Applying these insights to other scale free networks may explain the ubiquity of heavy-tailed degree distributions as the predictable result of Bernoulli Processes with high variance node fitness, particularly in bipartite graphs like collaboration networks.


\bibliography{Bipartite}

\end{document}